# Completing the Census of Exoplanets with the Microlensing Planet Finder (MPF)

RFI Response for the Astro2010 Program Prioritization Panel,

(The Basis for the Exoplanet Program of the WFIRST Mission)

David P. Bennett[1]

(phone: 574-631-8298, email: bennett@nd.edu),

J. Anderson[2], J.-P. Beaulieu[3], I. Bond[4], E. Cheng[5], K. Cook[6], S. Friedman[2], B.S. Gaudi[7], A. Gould[7], J. Jenkins[8], R. Kimble[9], D. Lin[10], J. Mather[9], M. Rich[11], K. Sahu[2], M. Shao[12], T. Sumi[13], D. Tenerelli[14], A. Udalski[15], and P. Yock[16]

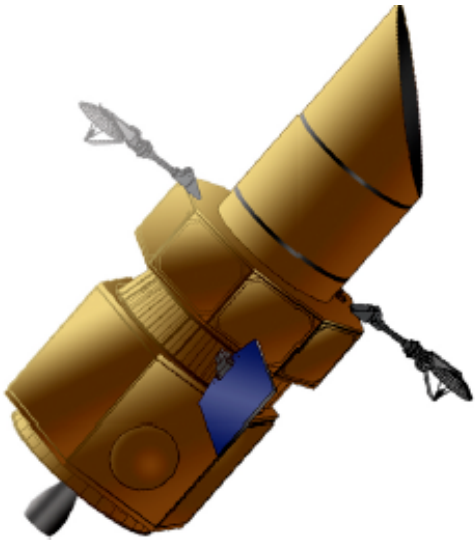
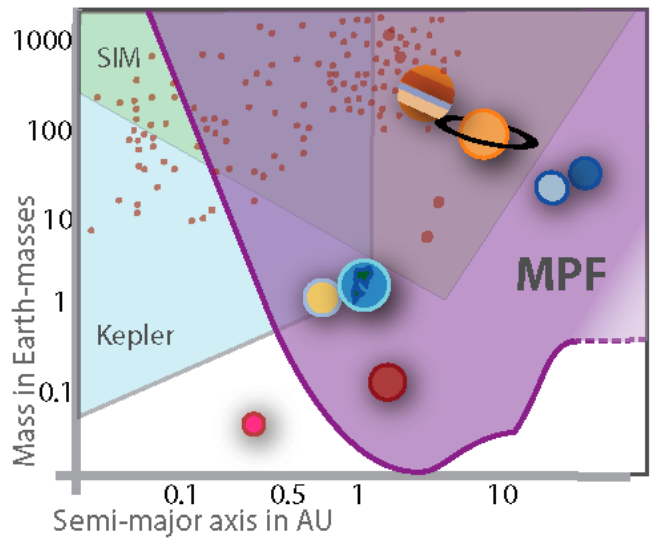

[1] University of Notre Dame, Notre Dame, IN, USA
[2] Space Telescope Science Institute, Baltimore, MD, USA
[3] Institut d'Astrophysique, Paris, France
[4] Massey University, Auckland, New Zealand
[5] Conceptual Analytics, LLC, Glen Dale, MD, USA
[6] Lawrence Livermore National Laboratory, USA
[7] Ohio State University, Columbus, OH, USA
[8] SETI Institute, Mountain View, CA, USA
[9] NASA/Goddard Space Flight Center, Greenbelt, MD, USA
[10] University of California, Santa Cruz, CA, USA
[11] University of California, Los Angeles, CA, USA
[12] Jet Propulsion Laboratory, Pasadena, CA, USA
[13] Nagoya University, Nagoya, Japan
[14] Lockheed Martin Space Systems Co., Sunnyvale, CA, USA
[15] Warsaw University, Warsaw, Poland
[16] University of Auckland, Auckland, New Zealand

The University of Notre Dame and its principal industrial partner, Lockheed Martin Space Systems Company, are pleased to provide this response to the Astro2010 Program Prioritization Panel. MPF is implememted using LM's low cost, low mass, spacecraft populated with flight proven components from Spitzer, XSS-11 missions and used in the IRIS program, and a proven telescope, lightweight mirror, detector designs; and line-of-sight stabilization technologies yielding a low risk program.

# 1. MPF Science Goals

MPF provides a statistical census of exoplanets with masses $\geq 0.1 M_\oplus$ and orbital separations ranging from 0.5AU to $\infty$. This includes analogs to all the Solar System's planets except for Mercury, as well as most types of planets predicted by planet formation theories. MPF measures the frequency of planets orbiting all types of stars with spectral type G or later. MPF and Kepler complement each other, and together they cover the entire planet-discovery space. Whereas Kepler is sensitive to close-in planets but is unable to sense the more distant ones, MPF is less sensitive to close-in planets, but surveys beyond 0.5 AU better than Kepler. MPF's sensitivity extends out even to unbound planets, offering the only possibility to constrain their numbers and masses. Other methods, including ground-based microlensing, cannot approach the sensitivity and comprehensive statistics on the mass and semi-major-axis distribution of extrasolar planets that a space-based microlensing mission provides. Thus, MPF provides the only way to complete the exoplanet census begun by Kepler and gain a comprehensive understanding of the architecture of planetary systems, needed to understand planet formation and habitability. MPF accomplishes these objectives with proven technology and a cost of under $330 million (FY 2009, excluding launch vehicle).

## 1.1. Basics of Gravitational Microlensing

The physical basis of microlensing is the gravitational bending of light rays by a star or planet. As illustrated in Fig. 1, if a "lens star" passes close to the line of sight to a more distant source star, the gravitational field of the lens star deflects the light rays from the source star. The gravitational bending effect of the lens star "splits", distorts, and magnifies the images of the source star. For Galactic microlensing, the image separation is ≤ 4 mas, so the observer sees a microlensing event as a transient brightening of the source as the lens star's proper motion moves it across the line of sight.

Gravitational microlensing events are characterized by the Einstein ring radius,

$$R_E = 2.0 \text{ AU} \sqrt{\frac{M_L}{0.5 M_e} \frac{D_L(D_S - D_L)}{D_S(1 \text{ kpc})}},$$

where $M_L$ is the lens star mass, and $D_L$ and $D_S$ are the distances to the lens and source, respectively. $R_E$ is the radius of the ring image that is seen with perfect alignment between the lens and source stars. The lensing magnification is determined by the alignment of the lens and source stars measured in units of $R_E$, so even low-mass lenses can give rise to high magnification microlensing events. A microlensing event's duration is given by the Einstein ring crossing time, typically 1-3 months for stellar lenses and a few days or less for a planet.

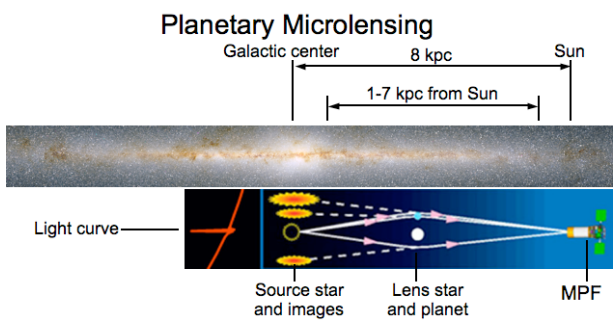

*Fig. 1: The geometry of a microlensing planet search towards the Galactic bulge. Main sequence stars in the bulge are monitored for magnification due to gravitational lensing by foreground stars and planets in the Galactic disk and bulge.*

**Planets are detected via light curve deviations** that differ from the normal stellar lens light curves (Mao & Paczynski 1991). Usually, the signal occurs when one of the two images due to lensing by the host star passes close to the location of the planet, as indicated in Fig. 1 (Gould & Loeb 1992), but planets are also detected at very high magnification where the gravitational field of the planet destroys the symmetry of the Einstein ring (Griest & Safizadeh 1998).

## 1.2. Capabilities of MPF

**MPF detects planets down to one tenth of an Earth mass**. The probability of a detectable planetary signal and its duration both scale as $R_E \sim M_p^{1/2}$, but given the optimum alignment, planetary signals from low-mass planets is quite strong. The limiting mass for the microlensing method occurs when the planetary Einstein radius becomes smaller than the projected radius of the source star (Bennett & Rhie 1996). The ~5.5 $M_\oplus$ planet detected by Beaulieu et al. (2006) is near this limit for a giant source star, but most microlensing events have G or K-dwarf source stars with radii that are at least 10 times smaller than this. So, the sensitivity of the microlensing method extends down to < $0.1 M_\oplus$, as the results of a detailed simulation of the MPF mission (Bennett & Rhie 2002) show in Fig. 2.

**Microlensing is sensitive to a wide range of planet-star separations and host star types.** The host stars for planets detected by microlensing are a random sample of stars that happen to pass close to the line-of-sight to the source stars in the Galactic bulge, so all common types of stars are surveyed, including G, K, and M-dwarfs, as well as white dwarfs and brown dwarfs. Microlensing is most sensitive to planets at a separation of ~$R_E$ (usually 2-3 AU) due to the strong stellar lens magnification at this separation, but the sensitivity extends to arbitrarily large separations. It is only planets well inside $R_E$ that are missed because the stellar lens images that would be distorted by these inner planets have very low magnifications and a very small contribution to the total brightness. These features are seen in Fig. 2, which compares the sensitivity of a space microlensing mission (MPF) with expectations for other planned and current programs. Other ongoing and planned programs can detect, at most, analogs of two of the Solar System's planets, while a space-based microlensing survey can detect seven—all but Mercury. The only method with comparable sensitivity to MPF is the Kepler space-based transit survey, which complements the microlensing method with sensitivity at semi-major axes, $a \leq 1$ AU. The sensitivities of MPF and Kepler overlap at separations of ~1 AU, which corresponds to the habitable zone for G and K stars.

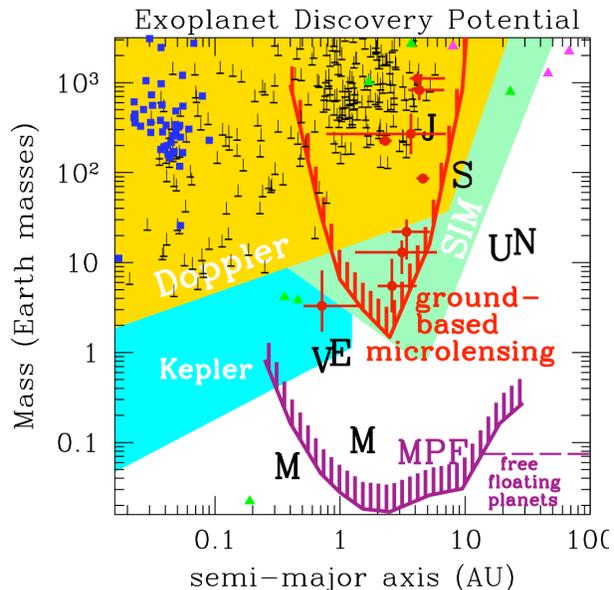

***Fig. 2:*** *Space-based microlensing (MPF) is sensitive to planets above the purple curve in the mass vs. semi-major axis plane. The gold, green and cyan regions indicate the sensitivities of radial velocity surveys, SIM and Kepler, respectively. Our Solar System's planets are indicated by their fist initials, and the known extrasolar planets are shown. Ground-based microlensing discoveries are in red; Doppler detections are inverted T's; transit detections are blue squares; timing and imaging detections are green and magenta triangles, respectively.*

The red crosses in Fig. 2 indicate the 5 gas giant (Bond et al. 2004; Udalski et al. 2005; Gaudi et al. 2008; Dong et al. 2008) and four $\sim 10 M_\oplus$ "super-earth" planets discovered by ground-based microlensing (Beaulieu et al. 2006; Gould et al. 2006; Bennett et al. 2008; Sumi et al. in preparation 2009). A preliminary analysis suggests that about one third of all stars are likely to have a super-earth at 1.5-4AU whereas radial velocity surveys find that only about 3% of stars have gas giants in this region (Butler et al. 2006).

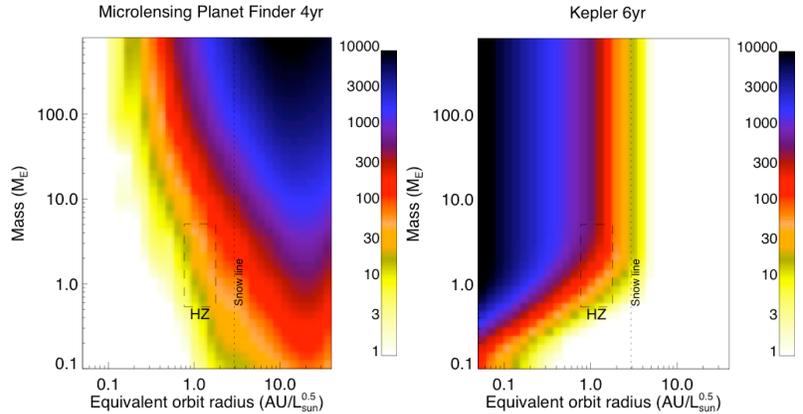

**Fig. 3:** MPF vs Kepler exoplanet "depth of survey" estimates from the Exoplanet Task Force report. The shading indicates the number of stars that are effectively searched for exoplanets as a function of mass and orbital radius, scaled by stellar luminosity so that the HZs of all types of stars are at ~ 1 AU.

Fig. 3 shows a comparison of the depth or survey estimates made by the Exoplanet Task Force, and this shows that a 4-year space microlensing mission finds a similar number of habitable zone (HZ) planets as an extended Kepler mission. Almost all the HZ planets found by MPF orbit G and K stars.

**Detailed simulations indicate a large number of planet detections.** Bennett & Rhie (2002) and Gaudi (unpublished) have independently simulated space-based microlensing surveys. These simulations included variations in the assumed mission capabilities that allow us to explore how changes in the mission design affect the scientific output, and they form the basis of our predictions in Figs. 2-5. Fig. 4 shows the expected number of planets that MPF would detect at orbital separations of 0.5-1.5, 1.5-5, and 5-15 AU. This plot assumes an average of one planet per star in each range of separations. Only MPF has a significant detection rate for Earths in all 3 separation ranges, and so only MPF can sample a wide range of separations including the HZ (see Fig. 3). Fig. 5 shows the expected detection rate for free-floating planets assuming one such planet per star. Free-floating planets are expected to be a common by-product of most planet formation scenarios, and only MPF detects free-floating planets of ≤ $1M_\oplus$, which are evidence of important dynamical interactions in the later stages of the planet formation process.

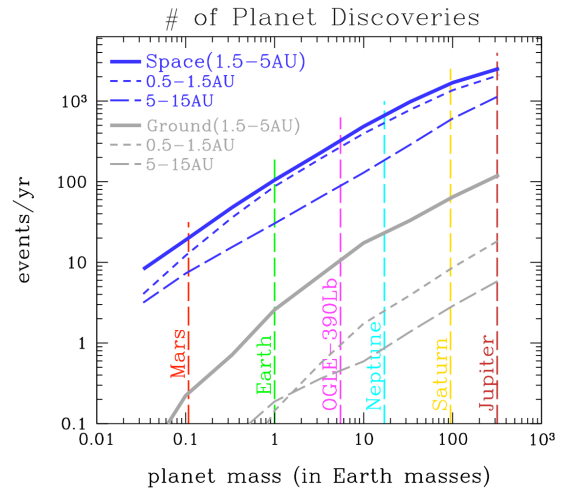

**Fig. 4:** The expected number of MPF planet discoveries as a function of the planet mass if every star has a single planet in the given separation of ranges.

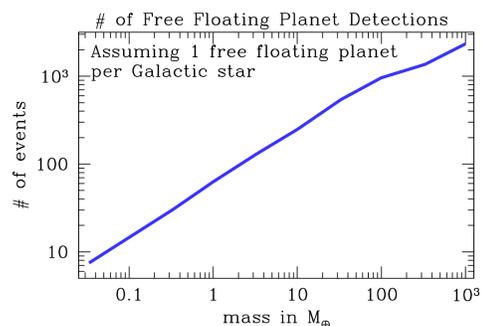

**Fig. 5:** The expected number of MPF free-floating planet discoveries.

### 1.3. MPF's Exoplanet Survey Is Needed

Microlensing relies upon the high density of source and lens stars towards the Galactic bulge to generate the stellar alignments needed to generate microlensing events, but this high star density also means that the bulge main sequence source stars are not generally resolved in ground-based images, as Fig. 6 demonstrates. This means that the precise photometry needed to detect planets of ≤ $1M_\oplus$ is not possible from the ground unless the magnification due to the stellar lens is moderately high. This, in turn, implies that ground-based microlensing is only sensitive to terrestrial planets located close to the Einstein ring (at ~2-3 AU). The full sensitivity to terrestrial planets in all orbits from 0.5 AU to ∞ comes only from a space-based survey.

**MPF light curves yield unambiguous planet mass ratios and separations.** For the great majority of events, the basic planet parameters (planet:star mass ratio, planet-star separation) can be "read off" the planetary deviation (Gould & Loeb 1992; Bennett & Rhie 1996; Wambsganss 1997). Possible ambiguities in the interpretation of planetary microlensing events have been studied in detail (Gaudi & Gould 1997; Gaudi 1998), and these can be resolved with high quality, continuous light curves that will be routinely acquired with a space-based microlensing survey.

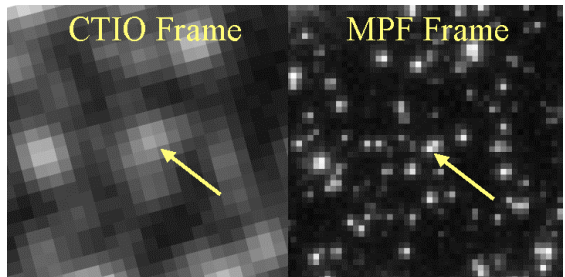

*Fig. 6: A comparison between an image of the same star field in the Galactic bulge from CTIO in 1" seeing and a simulated MPF frame (based on an HST image). The indicated star is a microlensed main sequence source star.*

**Planetary host star detection from space yields precise star and planet parameters.** For all but a small fraction of planetary microlensing events, space-based imaging is needed to detect the planetary host stars, and the detection of the host stars allows the star and planet masses and separation in physical units to be determined (Bennett et al. 2007). This can be accomplished with HST observations for a small number of planetary microlensing events (Bennett et al. 2006), but space-based survey data is needed for the detection of host stars for the thousands of planetary microlensing events that we expect from a space mission. Fig. 7 shows the distribution of planetary host star masses and the predicted uncertainties in the masses and separation of the planets and their host stars (Bennett et al. 2007) from simulations of the MPF mission. The host stars with masses determined to better than 20% are indicated by the red histogram in Fig. 7(a), and these are primarily the host stars that can be detected in MPF images. The distance to the planetary system is determined when the host star is identified, so a space-based microlensing survey will also measure how the properties of exoplanet systems change as a function of distance from the Galactic Center.

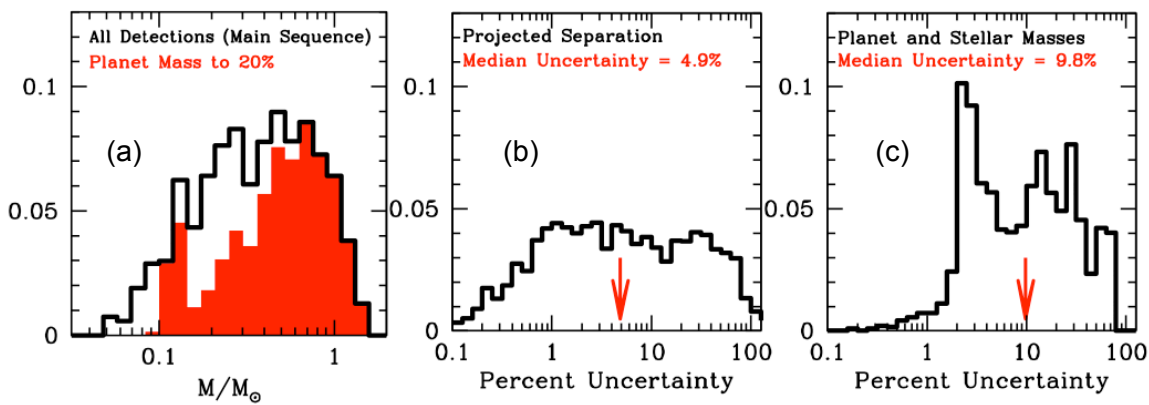

*Fig. 7: (a) The simulated distribution of stellar masses for stars with detected terrestrial planets. The red histogram indicates the subset of this distribution for which the masses can be determined to better than 20%. (b) The distribution of uncertainties in the projected star-planet separation. (c) The distribution of uncertainties in the star and planet masses.*

**Data analysis methods have been proven with ground-based and HST data.** HST transit and ground-based microlensing surveys have demonstrated the photometric precision needed by MPF, and current light curve modeling methods need only be automated for MPF.

### 1.4. MPF Constrains Planet Formation Theories

Rapid advancement in exoplanet research is driven by both extensive observational searches around mature stars as well as the construction of planet formation and models. Perhaps the most surprising discovery so far is the great diversity in the planets' dynamical properties, but these results are largely confined to planets that are unusually massive or reside in very close orbits. The core accretion theory suggests most planets are much less massive than gas giants and that the critical region for understanding planet formation is the "snow-line", located in the region (1.5-4 AU) of greatest microlensing sensitivity (Ida & Lin 2005; Kennedy et al. 2006). Early results from ground-based microlensing searches (Beaulieu et al. 2006; Gould et al. 2006; Bennett et al. 2008) appear to confirm these expectations. A space-based microlensing survey would extend the current sensitivity of the microlensing method down to masses of ~$0.1 M_\oplus$ over a large range (0.5AU-∞) in separation, and in combination with Kepler, such a mission provides sensitivity to sub-Earth mass planets at all separations. The semi-major axis region probed by space-microlensing provides a cleaner test of planet formation theories than the close-in planets detected by other methods, because planets discovered at > 0.5 AU are more likely to have formed *in situ* than the close-in planets. The sensitivity region for space-microlensing includes the outer habitable zone for G and K stars through the "snow-line" and beyond, and the lower sensitivity limit reaches the regime of planetary embryos at ~$0.1 M_\oplus$. Perhaps such planets are much more common than planets of $1 M_\oplus$ because their type-1 migration time is much longer.

**The habitability of a planet depends on its formation history.** The suitability of a planet for life depends on a number of factors, such as the average surface temperature, which determines if the planet resides in the habitable zone. However, there are many other factors that also may be important, such as the presence of sufficient water and other volatile compounds necessary for life (Raymond et al. 2004; Lissauer 2007). Thus, a reasonable understanding of planet formation is an important foundation for the search for nearby habitable planets and life.

### 1.5. Exoplanet Task Force Endorsement

The Exoplanet Task Force (ExoPTF) recently released a report (Lunine et al. 2008) that evaluated all of the current and proposed methods to find and study exoplanets, and they expressed strong support for space-based microlensing. Their finding regarding space-based microlensing states that: *"Space-based microlensing is the optimal approach to providing a true statistical census of planetary systems in the Galaxy, over a range of likely semi-major axes, and can likely be conducted with a Discovery-class mission."* Their conclusion that a space-based microlensing survey can be conducted with a Discovery-class mission, is in agreement with the judgment of the 2006 Discovery review panel, which found that the proposed Microlensing Planet Finder costs were credible. The ExoPTF's support of space-based microlensing is also reflected in their recommendation B.II.2, which states: "*Without impacting the launch schedule of the astrometric mission cited above, launch a Discovery-class space-based microlensing mission to determine the statistics of planetary mass and the separation of planets from their host stars as a function of stellar type and location in the galaxy, and to derive $\eta_\oplus$ over a very large sample.*"

## 2. MPF Technical Overview

The MPF uses a small telescope with a wide field to monitor $2\times10^8$ stars in the Galactic bulge for 9 months per year. For three months per year, the Sun is near the Bulge and MPF is available for other science. The MPF flies in an inclined geosynchronous orbit for continuous ground contact. Thick shields protect the detector from trapped cosmic rays. Normal operations are a repeated pattern of observations of several adjacent fields, combined with small dithers for improved spatial resolution. Twice a year, the MPF rolls 180 degrees to keep the Sun out of the instrument. The spacecraft uses avionics flown on Spitzer, XSS-11 and the up-coming IRIS mission, and achieves fine pointing with sensors in the main focal plane.

| Parameter | Value |
|---|---|
| Configuration | Folded three-mirror anastigmat, all conic mirrors |
| Aperture | 1.1 m |
| Field of View | 0.95°x0.68° |
| Effective Focal Length | 15.5 m |
| Detector Sampling | 0.24 arc-sec |
| Ensquared Energy | > 40% |
| Diffraction limit | < 1 μm |
| Spectral Range | 600-1700 nm with 3 selectable filters |
| Mirror Coating | Protected Silver |
| Operating Temperature | 270 K, actively controlled |
| Sun Avoidance Angle | > 45° |

*Table 1*: OTA Design Parameters

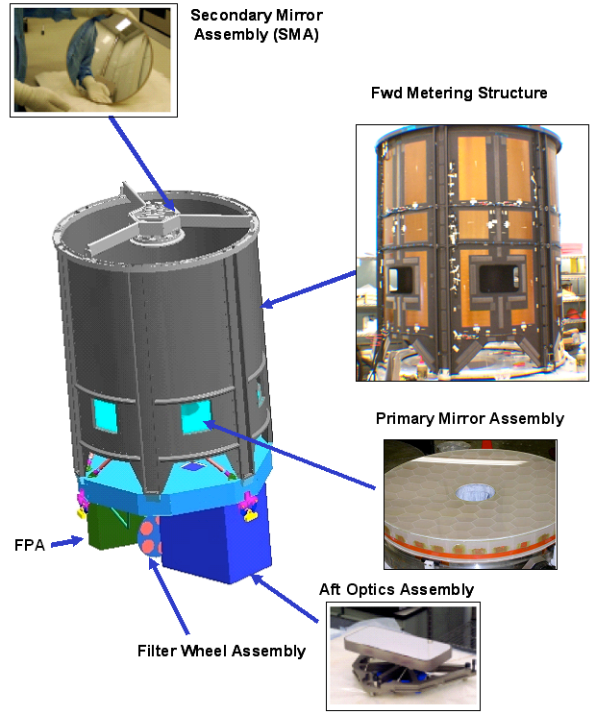

*Fig. 8: OTA Components*

The MPF observatory total (dry) mass is 820 kg and its total required power is 445W. Both of these numbers include contingency. The MPF mission is compatible with the Delta 2920-9.5, Delta 4040-12, or Atlas V 501 launch vehicles.

### 2.1. Science Instrument

The instrument consists of the Optical Telescope Assembly (OTA), based on a three-mirror anastigmat optical design, and the Focal Plane Array (FPA) with associated electronics.

### 2.1.1. OTA Requirements and Performance

The major OTA components are shown in Figure 8, with parameters given in Table 1. Starlight entering the OTA is collected by the Primary Mirror Assembly (PMA) that reflects the image to the Secondary Mirror Assembly (SMA), which reflects the image back through the center hole of the Primary Mirror (PM) into the Aft Optics Assembly (AOA) where the image is redirected and focused onto the FPA.

The SMA contains redundant actuators for focus adjustment. The AOA contains the Fold

Mirror Assembly (FMA), Tertiary Mirror Assembly (TMA), Filter Wheel Assembly (FWA) and the Radiation Lens Assembly (RLA). The FWA at the exit pupil selects three passbands.

The PM, SM, and TM optics are made of lightweighted ULE™ glass. The OTA support structures are made from low CTE composite laminates. Three kinematic mount flexures support the OTA. A block diagram is shown in Figure 9.

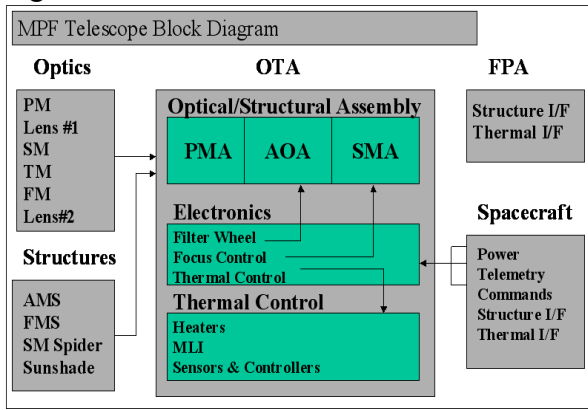

*Fig. 9: OTA Block Diagram*

The focal plane is contained within the low-temperature controlled FPA shroud, a passively controlled thermal enclosure serving as a stray-light baffle and a radiation shield. The FPA is thermally isolated and kinematically mounted to the telescope. To protect the detectors from cosmic rays, they are located in a shielded chamber following a filter wheel and a radiation protection lens located near the exit pupil, as shown in Figure 10.

The observatory sunshade protects the OTA from sunlight, and several internal baffles

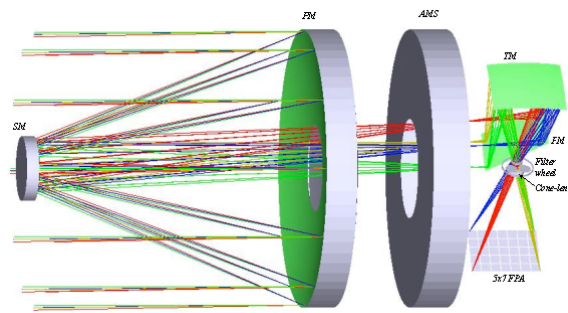

*Fig. 10: OTA Optical Configuration*

eliminate stray light within the OTA.

### 2.1.2. Focal Plane Array (FPA)

The MPF detector is a 146.8 megapixel, passively cooled Focal Plane Array (FPA), mounted on the OTA, with a Focal Plane Electronics (FPE) unit in the S/C bus. They provide fine guidance updates to the S/C from a guide star window in each SCA. Key parameters are shown in Table 2. Note that these are minimal performance requirements and that the FPA performance is expected to be much better, providing margin for the mission's scientific success.

**FPA Design**

As shown in Figure 11, individual SCA units are mounted on a base plate, electrically connected to PC boards under the base plate. Each board contains a SIDECAR ASIC providing clock and bias signals to drive 5 SCAs and digitize the images.

| FPA Parameter | Performance |
|---|---|
| Band | 600 - 1700 nm |
| Mean QE from 700 - 1600 nm | 55% |
| from 900 - 1400 nm | 75-80% |
| QE Uniformity | < 20% of Mean |
| Operating Temperature | 140 +/- 2K |
| Dark Current | < 1 e/sec (< 0.1 expected) |
| Mean Single Read Noise (CDS) | < 30 e |
| Pixel Size | 18 um |
| Array Format | 2048 x 2048 |
| # of SCAs | 35 (5x7 mosaic) |
| Operability | > 95% |
| Integration Time | 2.8 min |
| FPA    Size (mm) | 380 x 280 x 127 |
| Mass | 18.3 kg |
| Power | 0.57 W |
| FPA Passive Thermal Load | < 1.6 W |
| Image Plane Z-axis Tolerance | < 50 um |
| Radiation Hardness | > 80 kRads |
| FPE    Size (mm) | 200 x 190 x 33 |
| Mass | 1.7 kg |
| Power | 15.6 W |

*Table 2: FPA and FPE Parameters*

Each detector is bump-bonded to a HAWAII-2RG readout integrated circuit (ROIC), featuring readout mode versatility,

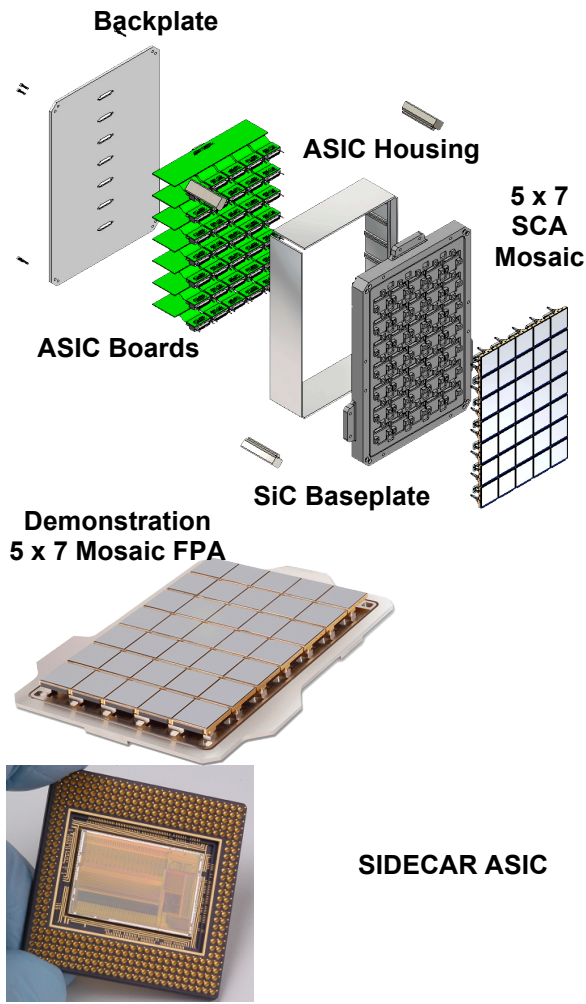

*Fig. 11: FPA Components*

guide star capability, zero glow, and low noise operation. The ROIC design already includes a high-speed guide star readout mode.

**Detector Producibility**

The initial H2RG detectors were produced over 8 years ago, and the current process is mature. The interconnect yield of 2Kx2K HAWAII-2 FPAs over the past 2 years is > 99.9%. Team member Lockheed Martin developed the Hubble Space Telescope (HST) and NICMOS. Team members developed Mercury-Cadmium-Telluride detectors (2.5 μm cutoff, liquid phase epitaxy), the HST Wide-Field Camera 3 (WFC3) detector (1.7 μm cutoff, molecular beam epitaxy), culminating in state-of-the-art performance for the James Webb Space Telescope (JWST) 2.5 μm and 5.0 μm cutoff arrays. Detectors with the same technology were used in NASA's Moon Mineralogy Mapper, the WISE all-sky infrared survey, and are flying to Pluto in the New Horizons' Ralph instrument.

The performance delivered by the WFC3 and JWST detectors is easily sufficient for the relatively bright objects that will be observed by MPF. The large number of detectors in MPF, coupled with its survey-mode strategy, substantially reduces the operability requirements on each detector. While the number of detectors required is relatively large, the requirements are loose and the yield is high.

Using the TIS SIDECAR ASIC for detector control and readout provides significant reduction in system-level complexity and cost. This device has been qualified for flight on JWST, and was installed in the Advanced Camera for Surveys during the HST Servicing Mission 4 where it has been operational since mid-2009.

A model of an MPF-like focal plane (photo in Fig. 11) has demonstrated the packaging techniques and the integration and alignment processes. Similar ground-based arrays have been successful, e.g. the SiC focal plane for the WIYN One-Degree-Imager.

### 2.1.3. Instrument Thermal Control Concept

Payload electronics are mounted inside the S/C bus, and the OTA and FPA are thermally isolated. The OTA is protected by multilayer insulation and controlled by a multiple zone feedback controlled heater system. The FPA is coupled to a dedicated sun-opposing cryo-radiator assembly.

### 2.2. Spacecraft and Mission Implementation

The flight system uses an operational Spitzer-heritage S/C design, flight proven components, a proven telescope design, with lightweight mirror and detector designs, and robust line-of-sight stabilization technologies, resulting in a low risk program.

### 2.2.1. Mission Parameters

MPF will be launched into a 28.7° inclined circular geosynchronous orbit. This orbit permits continuous communication with a ground station at White Sands, NM, allowing real-time operations and continuous data down link. It also allows the MPF Galactic bulge field to be observed for 9 months per year, leaving 3 months per year available for a GO program.

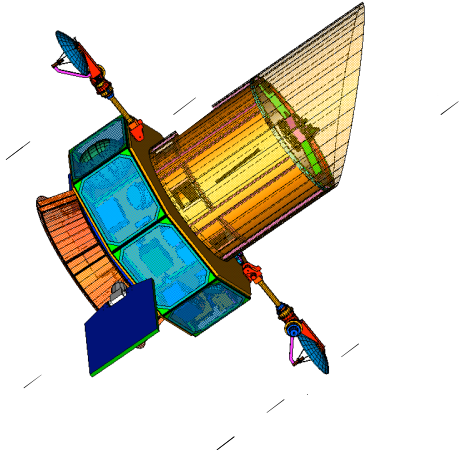

*Fig. 12: MPF Spacecraft Schematic*

### 2.2.2. Spacecraft Design Concept

The proposed S/C bus (TRL 8.5) provides <1 arcsec stability over 1 day using bus-mounted star trackers and gyros. Figure 12 shows the MPF spacecraft.

**Propulsion Concept**

The MPF propulsion subsystem uses 4 hydrazine thrusters for pitch/yaw attitude control and 4 hydrazine Roll/fine orbit adjust thrusters). Sufficient propellant is provided to perform orbit and momentum management for 5 years, and dispose of the observatory at end of life.

**Structure Concept**

The modular design has 9 bays, with four side modules, four corner modules, two heat-pipe equipment panels, and six equipment panels. The equipment panels are sandwiched between the modules.

**Mechanisms**

Gimbal mechanisms deploy and point the two HGAs (high-gain antennae) and both solar array wings.

**Pointing Control System (PCS) Concept**

The PCS utilizes proven software algorithms and redundant COTS hardware. The PCS modes are described in Table 3. The fine pointing mode architecture utilizes reaction wheels, gyros, and the Focal Plane Array.

Sun sensors, star trackers, and rate gyros are used to control attitude in transitional and contingency modes of operation and to orient the observatory within the acquisition range of the instrument's fine guidance function.

| Mode | Description |
|---|---|
| Orbit Injection | Acquire sun and maintain power positive |
| Coarse Pointing | Coarse attitude determination and stabilization |
| Slew | Large angle maneuvers of 0.7 and 1.4 deg |
| Fine Pointing | Fine pointing during science observation |
| Momentum Dump | Thruster burns dump wheel angular momentum |
| Safe Hold | Power positive sun sensor hold |

*Table 3: Pointing Control Concept*

To acquire the target region, the instrument focal plane is used as a star camera to match an on-board star catalog. Windows on the focal plane array track guide stars and provide fine guidance during observations. 3-axis attitude control is provided by four reaction wheel assemblies (RWAs), operated with zero net momentum, biased to avoid vibrational harmonics. RWA momentum is unloaded using thrusters, and the solar array and high gain antenna are moved, between sensitive observations.

The photometric pointing stability requirement of 0.2 pixel determines the needed pointing stability of 0.048 arcsec. Modeling shows that the S/C meets the requirements. The Kepler mission uses the same concept, deriving error signals from its large mosaic detector.

**Electrical Power Concept**

The power design includes flight-proven designs for the Solar Array, Battery, Charge Control Unit and Power Distribution and Drive Unit. Electrical power is generated by two Iridium-heritage rigid arrays mounted on the +$Y$ and -$Y$ sides of the bus.

**C&DH (Command and Data Handling) Concept**

The fully redundant C&DH subsystem operates the S/C with commands stored in memory or via real-time commands, as well as handling engineering and science data.

**Thermal Control Concept**

Heat pipes run the length of the bus, transferring heat to the bay sides with outboard radiators. The battery and rate gyro packages have dedicated radiators. Conduction and radiation, with thermostatically controlled heaters, control the internal equipment.

**Communications Concept**

The MPF telecomm subsystem provides reception and transmission of commands and telemetry data through an S-Band link. The science data are downlinked do a dedicated ground station using Ka-band.

## 3. Technology Drivers

MPF requires no technology development. All required technologies are at TRL 6 or higher, with the possible exception of the SiC baseplate for the FPA. The basic technology for the SiC baseplate is at TRL 6, and the specific implementation for MPF will be brought to TRL 6 in phase A.

## 4. Organization and Current Status

### 4.1. MPF Organization

The MPF team is led by Principal Investigator, David Bennett, who is a pioneer in the gravitational microlensing field, having helped to initiate the MACHO Project, which discovered the first gravitational microlensing event in 1993. He was (with S.H. Rhie) the first to show that gravitational microlensing could detect Earth-mass planets, and he has played a lead role in the analysis of most of the 11 exoplanets found by microlensing to date. These discoveries include two that were the lowest mass exoplanet known at the time of their discovery, including the current record holder, MOA-2007-BLG-192Lb (Bennett et al. 2008).

Because there is very little space hardware experience in the gravitational microlensing community, the role of Deputy Principal Investigator, Edward Cheng, is critically important. Cheng has 20 yrs experience in detectors and spaceflight missions. He has served as Deputy Project Scientist for the Cosmic Background Explorer (COBE), Manager of the NASA/GSFC Detector Characterization Laboratory Instrument Scientist for the Hubble Space Telescope (HST) Wide Field Camera 3, Chief Scientist for the HST NICMOS Cooling system, and Principal Investigator for the HST Advanced Camera for Surveys repair mission.

MPF is managed by the NASA/Goddard Space Flight Center, with the help of MPF's Goddard Science Team Leader, John Mather. Mather is NASA's only Nobel Prize winner, and he will recruit some of Goddard's top engineers to work on MPF, when it is selected for flight.

MPF's spacecraft will be provided by Lockheed Martin and the telescope and focal plane will be provided by ITT and Teledyne Imaging Sensors, respectively.

A group from the Space Telescope Science Institute including Scott Friedman and Jay Anderson will handle the primary data analysis and archiving, and a group under the direction of the PI at the University of Notre Dame will do the light curve analysis.

### 4.2. Current Status

MPF was proposed to the 2004 and 2006 NASA Discovery competitions, and it received a top science rating in the 2004 competition, and an acceptable rating in the technical review of the 2006 competition. To our knowledge, MPF was the only exoplanet mission proposal

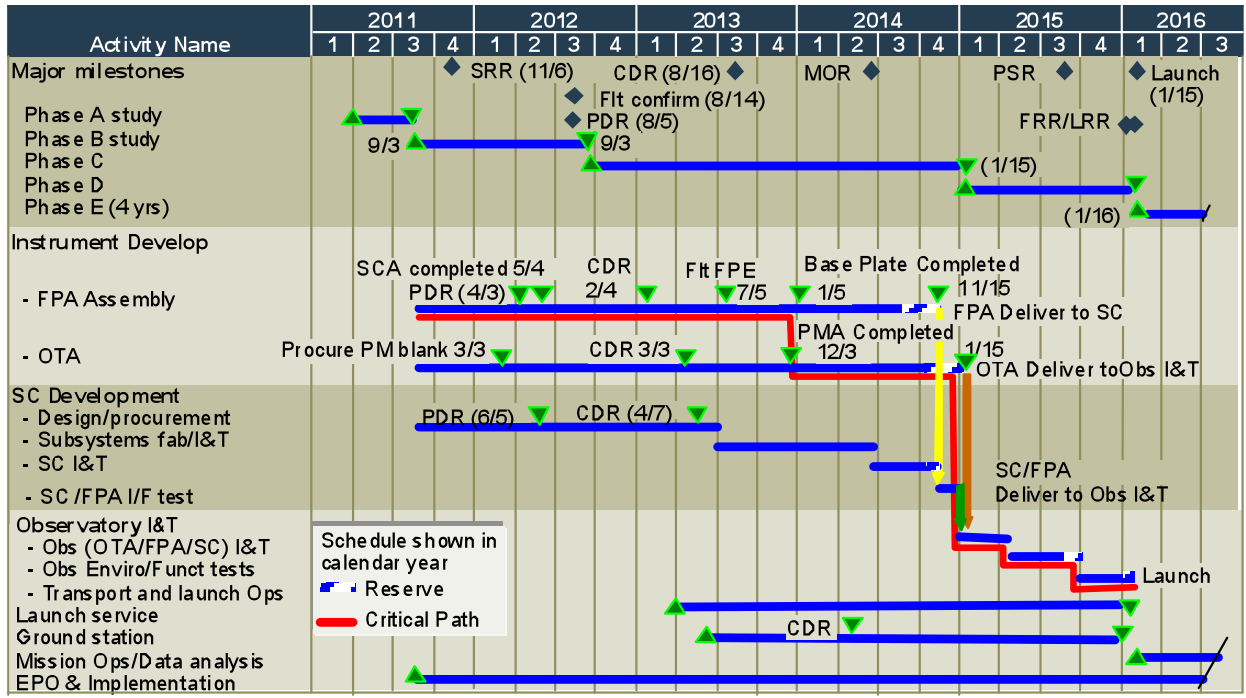

***Fig. 13:*** *The MPF mission schedule with funded reserve and critical path analysis*

that the 2006 Discovery technical review did not rate as high risk.

Since the 2006 Discovery competition a number of MPF components, such as the H2RG detectors and the SIDECAR ASIC electronics have advanced in their development, so MPF's technology is significantly more mature now than it was in 2006.

Most recently, a space-based microlensing planet survey, like MPF, was considered and endorse by last year's Exoplanet Task Force (Lunine et al. 2008), provided that it could be done (like MPF) at the cost of a Discovery mission. However, NASA has recently decided to remove exoplanet missions from the Discovery line in favor of a new Exoplanet Probe line. If this new Exoplanet Probe line is started, it is expected to have a significantly higher cost cap than Discovery. So, MPF would be much cheaper than the Exoplanet Probe cost cap.

## 5. MPF Schedule

Figure 13 is a summary of the MPF schedule for a 4-year mission, showing the critical path in red. This schedule follows the constraints of the NASA Discovery program, which usually limits phase C/D to 4 years or less. This is appropriate for low-cost missions, like MPF, which do not require technology development.

The schedule has more than 30 days per year (Phase B-D) of funded reserve. Trade studies, SSR, PDR, and procurement of long lead items are scheduled for Phase B. Phase C/D contains the CDR, flight build, integration, environmental test campaign, launch on 15-Jan-2016, and a 30-day initial on-orbit checkout (IOC). Phase E, the science operations and data analysis phase, starts at the end of IOC and lasts 4 years, ending on 15-Feb-2020.

## 6. MPF Cost Estimates

The MPF mission cost was studied quite thoroughly during the preparation of the MPF proposal that was submitted to the 2006 NASA Discovery competition. The cost studies included a bottoms-up analysis by the MPF team, including NASA/GSFC, Lockheed Mar-tin, ITT, Teledyne Imaging Sensors, the Space Telescope Science Institute, and the University of Notre Dame. The only significant change for the costs presented here is that the costs were escalated for inflation from FY 2006 to FY 2009. The costs are summarized in Table 4.

The use of the 2006 cost estimates is conservative because many of MPF's technologies have become more mature during the past three years.

| Phase | Cost (FY 2009 k$) |
|---|---|
| A | $ 1,305 |
| B | $ 39,879 |
| C/D | $ 259,213 |
| E | $ 28,427 |
| **Total NASA Cost** | **$ 328,824** |
| Contributions | $ 4,385 |
| **Total Price (excluding ELV)** | **$ 333,209** |

*Table 4: MPF Costs*

### 6.1. MPF Cost Models

In addition to this bottoms-up analysis, costs were also estimated using two commercially available hardware cost models, PRICE H and SEER-H. These analyses were done in March, 2006, by the GSFC Resource Analysis Office (RAO).

Both the PRICE H and SEER-H cost modeling programs produced cost estimates that came within 10% of the bottoms up costs. Similarly, the PRICE H and SEER-H models estimated spacecraft costs that were very close to the bottoms-up cost estimate.

At the time that the SEER-H analysis was run, the model was ill equipped to handle the costing of the IR detectors that will be used for MPF. Because of this difficulty with SEER-H, we used the actual costs for the JWST NIRSpec detectors in lieu of SEER-H to estimate the detector costs. This is a conservative procedure because the JWST NIRSpec detector requirements are much more stringent than the MPF requirements, and the production process has continued to improve at Teledyne Imaging Sensors. So, it is very likely that the MPF detectors can be produced with a higher yield of flight-quality detectors.

### 6.2. 2006 Discovery Cost Review

The MPF cost estimates presented here were reviewed by the TMCO (Technical, Management, Cost, Other) panel for the 2006 NASA Discovery program review. The TMCO panel did three independent cost estimates, and then compared them to reach a consensus cost. Their consensus cost was within 12% of the costs proposed by the MPF team and presented here.

### 6.3. Post-2006 Technology Development

The technology involved with several of the MPF components has matured significantly since the MPF Discovery proposal was submitted. The TIS HgCdTe detectors and SIDECAR ASIC electronics chips have been delivered for JWST and are fully flight qualified. Similarly, 1K×1K TIS HgCdTe detectors with the same 1.7μm long wavelength cutoff used for MPF have been installed on HST as part of the Wide Field Planterary Camera 3 during the HST Servicing Mission 4 in May 2009, and substrate-removed HgCdTe sensors from Teledyne have contributed to the success of NASA's WISE and Moon Mineralogy Mapper instruments. The SIDECAR ASIC is the heart of an electronics module that replace the failed electronics unit for the HST Advanced Camera for Surveys, and the SIDECAR ASIC has been delivered to NASA for use in the Landsat Data Continuity Mission Thermal InfraRed Sensor (LDCM TIRS).

The technology behind the MPF OTA has also matured with the successful launch of GeoEye I (2008) and WordView II (2009) Earth-observing imaging payloads (1.1m aperture) developed by ITT under the National Geospatial Intelligence Agency (NGA) sponsored NextView satellite program. (See Figure 15.)

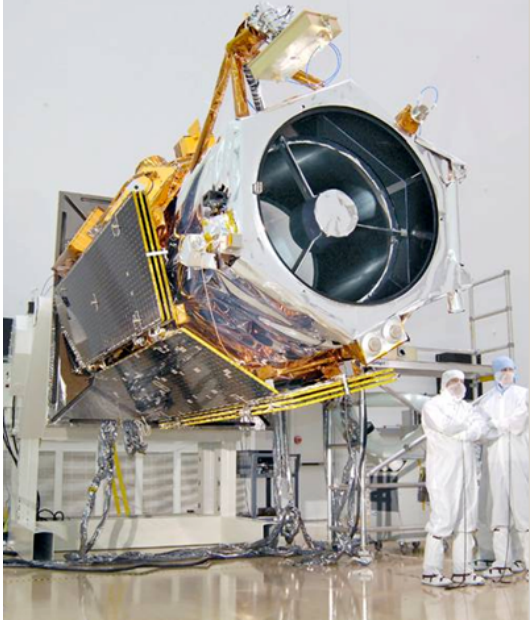

*Fig. 14*: Geoeye I Earth Imaging Payload

## 6.4. Optional General Observer Program

MPF can only search for planets in its Galactic bulge fields for 9 months per year when the bulge is more than 45° away from the sun. The other 3 months per year can be devoted to an optional General Observer (GO) program, but the costs of administering this GO program are not included in this cost estimate. However, mission operations and MPF pipeline data reduction during these 3 months per year are included.

## 7. MPF Summary

The Microlensing Planet Finder (MPF) is a low-cost planetary survey mission that will complete the census of extrasolar planetary systems that is now being begun by Kepler. While Kepler will provide important results on the frequency of close-in planets at separations ≤ 1 AU, MPF is needed to learn about planets in outer orbits. MPF is sensitive to planets down to 0.1 Earth-masses at orbital separations of ≥ 0.5 AU, including the habitable zone for G and K stars. The sensitivity region of MPF (see Fig. 2) includes 7 of the 8 solar system planets, as well as the vicinity of the "snow-line" where the first planets are predicted to form by the core accretion theory. This is also the region where the Earth's water is thought to originate, so an understanding of the planet formation process near the "snow-line" is needed for considerations of planetary habitability.

| Property | Value | Units |
|---|---|---|
| Orbit | Inclined GEO 28.7 | degrees |
| Mission Lifetime | 4.0 | years |
| Telescope Aperture | 1.1 | meters (diam.) |
| Field of View | 0.95 × 0.68 | degrees |
| Spatial Resolution | 0.240 | arcsec/pixel |
| Pointing Stability | 0.048 | arcsec |
| Focal Plane Format | 146.8 | Megapixels |
| Spectral Range | 600 – 1700 | nm in 3 bands |
| Quantum Efficiency | > 75% <br> > 55% | 900-1400 nm <br> 700-1600 nm |
| Dark Current | < 1 | e-/pixel/sec |
| Readout Noise | < 30 | e-/read |
| Photometric Accuracy | 1 or better | % at J = 20.5. |
| Data Rate | 50.1 | Mbits/sec |

*Table 5:* Key MPF Mission Requirements

The importance of MPF's census of extrasolar planets at orbital separations of ≥ 0.5 AU was recognized by the Exoplanet Task Force, which recommended a space-based microlensing survey if it could be done at the cost of a Discovery mission.

The MPF mission requirements are summarized in Table 5. MPF continuously observes four 0.65 sq. deg. fields in the central Galactic Bulge using an inclined geostationary orbit to provide a continuous view of the Galactic Bulge fields and a continuous downlink. MPF will use a dedicated ground station co-located with other NASA facilities at White Sands, NM. Spacecraft commanding and on-board processing are minimized because of the simple observation plan and orbit design.

**MPF system.** MPF uses a 1.1m Three-Mirror Anastigmat (TMA) telescope feeding a 146.8 Mpixel HgCdTe focal plane residing on a standard spacecraft. The MPF design leverages existing hardware and design concepts, many of which are already demonstrated on-orbit and/or flight qualified. The spacecraft bus is a near-identical copy of that used for Spitzer and has demonstrated performance that meets MPF requirements. The telescope system leverages Ikonos and NextView commercial Earth-observing telescope designs that provide extensive diffraction-limited images. The focal plane design taps proven technologies developed for JWST. All elements are at TRL 6 or better. The focal plane design uses common non-destructive readout CMOS multiplexers for two detector technologies that cover the visible and the near-IR. The MPF focal plane can track up to 35 guide stars in the field, providing a built-in fine guidance capability. The focal plane gains additional advantages from using the Teledyne SIDECAR™ application specific integrated circuit (ASIC) that condenses all the control and readout electronics into a "system-on-a-chip" implementation. This approach dramatically simplifies the support electronics while minimizing wire-count challenges.

**Cost.** The total cost for the MPF mission is (FY09) $330M including 30% contingency during development, but excluding the launch vehicle.